\documentclass[12pt,preprint]{aastex}

\slugcomment{Preprint DAMTP-2001-96; April 18th, 2002}
\shorttitle{Modified Median Statistics and Type Ia Supernova Data}
\shortauthors{P.P. Avelino, C.J.A.P. Martins and P. Pinto}

\begin{document}
\title{Modified Median Statistics and Type Ia Supernova Data}

\author{P.P. Avelino\altaffilmark{1}, C.J.A.P. Martins\altaffilmark{2,3}
and P. Pinto\altaffilmark{1}}
\affil{Centro de Astrof\'{\i}sica da Universidade do Porto,\\
Rua das Estrelas s/n, 4150-762 Porto, Portugal}

\altaffiltext{1}{Dep. de F{\' \i}sica da Faculdade de Ci\^encias da
Univ. do Porto, Rua do Campo Alegre 687, 4169-007 Porto, Portugal.
Email: pedro\,@\,astro.up.pt}
\altaffiltext{2}{Department of Applied Mathematics and Theoretical Physics,
Centre for Mathematical Sciences, University of Cambridge,
Wilberforce Road, Cambridge CB3 0WA, United Kingdom.
Email: C.J.A.P.Martins\,@\,damtp.cam.ac.uk}
\altaffiltext{3}{Institut d'Astrophysique de Paris,
98 bis Boulevard Arago, 75014 Paris, France}

\begin{abstract}
The median statistic has recently been discussed by Gott \textit{et al.}
as a more reliable alternative to the standard $\chi^2$ likelihood
analysis, in the sense of requiring fewer assumptions about the data
and being almost as constraining.
We apply this statistic to the currently available combined
dataset of 92 distant type Ia supernovae, and also to a mock SNAP-class
dataset. We find that the performances of the modified median and $\chi^2$
statistics are comparable, particularly in the latter case.
We further extend the work of Gott \textit{et al.} by modifying the median 
statistic to account for the number and size of sequences of consecutive 
points above or below the median. We also comment on how the performance 
of the statistic depends on the choice of free parameters that one is 
estimating.
\end{abstract}

\keywords{Cosmology; data analysis; statistical methods; type Ia supernovae}

\section{Introduction}

In recent work \citet{Gott:2000mv} have argued, through
several convincing examples, that the median statistic is a reliable
alternative to the usual $\chi^2$ likelihood analysis. Even though it
usually has the caveat of not being as constraining (for the same data set),
it has the strong advantage of requiring much weaker assumptions about the
dataset itself and its errors. Furthermore, it is also less vulnerable than
the mean to the presence of bad data, such as when `outliers' exist.
Hence, if nothing else, median statistics can be useful for
the early stages of a data analysis pipeline, when one is still trying
to put together evidence that may justify the use of stronger hypotheses
on the dataset.

In this work, we apply the median statistic to the combined dataset of
92 type Ia supernovae taken from the Supernova Cosmology Project
(SCP) \citep{Perlmutter:1998np} and the High-z Supernova Search
Team (HzST) \citep{Riess:1998cb}\footnote{Note that in
\protect \citep{Gott:2000mv} median statistics was only separately applied to
earlier versions of each dataset.}. This is a particularly relevant example,
since the current data set is still fairly small. Furthermore, the
physics of supernova explosions is not at all well known, and hence the
possible presence of outliers in the available data is a particular concern.

We also apply our results to a SNAP-class \citep{Nugent:2000vs,Weller:2001gf}
simulated dataset. We further extend the analysis of \citet{Gott:2000mv}
by considering simple modifications  of the median statistic which account
for the number and size of sequences of  consecutive points above or below
the median (to which the standard median statistic is `blind'). We discuss
the dependence of the performance of 
the statistic on the choice of free parameters that one is estimating and 
we find that the median statistic can be competitive with the standard
$\chi^2$ analysis, provided  one knows the Hubble parameter
$H_0$ (or equivalently the absolute magnitude of a standard type Ia 
supernova). 

The rest of the paper is organised as follows. In Section 2 we
briefly recall the tools necessary to carry out the estimation of the
present day  values of the cosmological parameters $\Omega_m$ and
$\Omega_\Lambda$ from a type Ia
supernova dataset. In Section 3 we introduce the median statistic
as a data analysis method and motivate some simple modifications thereof.
We then present our results in Section 4, and finally in
Section 5 we draw some conclusions and discuss possible further
improvements.

\section{Cosmological parameters from type Ia supernovae}

Following the release of the results from the Supernova Cosmology 
Project (SCP) \citep{Perlmutter:1998np} and the High-z Supernova Search Team 
(HzST) \citep{Riess:1998cb}, which altogether include some 100 supernovae,
there has been an increased effort towards the parameterisation of the energy
content of the Universe using Type Ia Supernovae \citep{Weller:2001gf}.

The currently available data, when combined with CMB 
results \citep{Jaffe:2000tx}
indicates that about one third of the critical energy density in the
Universe is in the form of ordinary matter (and here we include the classic
dark matter), while the remaining two thirds are in the form of a so-called
dark energy component, the exact form of which is yet
unknown. The cosmological constant $\Lambda$ is arguably the simplest
candidate for this dark energy \citep{Bean:2001xy}, though there are various
other contenders, from frustrated topological defects \citep{Bucher:1998mh}
to a time varying cosmological constant \citep{ratra1,ratra2,ratra3,waga},
in particular what is commonly called
quintessence \citep{Caldwell:1998ii,Wang:1999fa}.
The main problem associated with the cosmological
constant is that theoretical predictions of its value are many orders of
magnitude off from the experimental results \citep{carroll}. On the other hand,
quintessence may suffer from considerable `coincidence' problems.

Aiming to help settling the question of the constitution of dark energy, the
SNAP (Supernova Acceleration Probe) \citep{Nugent:2000vs}
satellite was recently proposed. Its goal is to obtain a supernova dataset
more than one order of magnitude larger than currently available datasets,
with much-improved control over systematic errors,
to redshifts up to about $z \simeq 1.7$.
Even though SNAP results are still years away, we can of course
simulate the expected results, and thus forecast the impact of this improved
dataset on the constraints on the energy density and equation of state of
dark energy that permeates the Universe \citep{Weller:2001gf}.

As usual, the parameter fit is done through the luminosity distance $d_L$,
defined as
\begin{equation}
{\mathcal F} = \frac{{\mathcal L}}{4\pi d_L{}^2}\,, \label{flux}
\end{equation}
where ${\mathcal L}$ is the intrinsic luminosity of the source and
${\mathcal F}$ the measured flux. From the Friedmann--Robertson--Walker
metric it follows that this distance is given,
as a function of redshift $z$, by
\begin{equation}
d_L = \frac{c(1 + z)}{H_0 \sqrt{|\Omega_k|}}\ S(\sqrt{|\Omega_k|} \int _0^z
(\Omega_m(z' + 1)^3 + \Omega_{\Lambda} + (z' + 1)^2
\Omega_k)^{-\frac{1}{2}}dz')\,, \label{ludist}
\end{equation}
where $\Omega_k = 1 - \Omega_m - \Omega_{\Lambda}$, and the function $S$
is defined as
\begin{equation}
S(x) = \left\{ 
\begin{array}{ll}
\sin x, & \Omega_k < 0 \\
x, & \Omega_k = 0 \\
\sinh x, & \Omega_k > 0
\end{array} \right.\,.\label{curvature}
\end{equation}

The apparent magnitude $m$ of a supernova (a parameter more often used than
the measured flux ${\mathcal F}$, to which it is related) at a given redshift
is then given by
\begin{equation}
m = M + 5\log \left(\frac{d_L}{{\rm {Mpc}}}\right) + 25\,, \label{mags}
\end{equation}
$M$ being the absolute magnitude of the supernova (related to its
intrinsic luminosity ${\mathcal L}$).
Following Wang \citep{Wang:1999bz}, we use results from both the SCP
and HzST even though their published datasets differ in presentation.
We define the distance modulus
\begin{equation}
\mu _0 = 5\log \left(\frac{d_L}{{\rm {Mpc}}}\right) + 25\,, \label{distmod}
\end{equation}
as presented in the HzST results comprising 50 supernovae. 
Comparatively, the SCP
published the estimated effective {\it B}-band magnitude
$m_b^{\textrm{eff}}$ for 60 supernovae which relates to the HzST results
through
\begin{equation}
\label{m-mu}
m_B^{\textrm{eff}} = M_B + \mu _0\,,
\end{equation}
where $M_B$ is the peak {\it B}-band absolute magnitude of a standard 
type Ia supernova.
The published results of the SCP and HzST groups have 18 common supernovae,
16 of which are from the Cal\'an--Tololo Survey data \citep{Phillips:1999vh}.
If we calculate $M_B$
by comparing results for these 18 supernovae (using the results from the HzST
estimated by means of the MLCS method) we get
\begin{equation}
M_B = m_B^{\textrm{eff}} - \mu _0 = -19.33\pm 0.25\,.\label{meff}
\end{equation}
Assuming the value $M_B = -19.33$ for the absolute luminosity, we convert the 
SCP results to distance modulus using eqn. (\ref{m-mu}). We then add 42 of
these supernovae to the dataset from HzST, leaving out the 18 already present,
thus making a total of 92 supernovas.

As for future results, our simulation assumes a specific set of parameters
based on the results from current available measurements
\begin{equation}
\Omega_m = 0.3,\quad \Omega_{\Lambda} = 0.7,\quad H_0 = 65.2\ {\rm km\
s^{-1}\ Mpc^{-1}}.
\end{equation}
We have simulated a supernova dataset with $\mu_0$ 
drawn from a Gaussian distribution with an
average computed from the above parameters and 
$\sigma_{\mu_0} = 0.15$ standard
deviation, divided in bins with similar characteristics to those of SNAP
projections, as described in Table \ref{tabsnap}.

\begin{table}
\caption{\label{tabsnap}Distribution of type Ia supernovae
by redshift bins in the mock SNAP-class simulated dataset.}
\begin{tabular}{|c|c|c|}
Min. redshift & Max. redshift & Number of SNe \\
\hline
0.0 & 0.2 & 50 \\
0.2 & 1.2 & 1800 \\
1.2 & 1.4 & 50 \\ 
1.4 & 1.7 & 15 \\
\end{tabular}
\end{table}

\section{Standard median statistics, and how to improve it}

Type Ia supernova data analysis is usually carried out using a $\chi^2$
analysis. Here, however, we shall describe median statistics as an alternative
analysis method, and then propose and motivate some simple
modifications of it and study the constraints which can be thus obtained.
Rigorous descriptions of the standard median statistic can be found in most 
good statistics textbooks. For a more detailed review emphasizing
some aspects relevant to astrophysics see \citep{Gott:2000mv}.

Recall that a $\chi^2$ statistical treatment requires that four hypothesis
be met, namely (1) that the experimental results are statistically independent;
(2) that there are no systematic errors present; (3) that the experimental
errors follow a Gaussian distribution; and (4) that the standard deviation of
these errors is known.

The fewer assumptions one needs to make about a given dataset,
the higher will be the confidence in the results derived from it. It turns out
that keeping only assumptions (1) and (2) and relaxing the others one
can still make quite strong statements.
Assuming that the experimental results are statistically independent and that
there were no systematic errors made, one expects that upon performing a
large number of measurements approximately half of the values obtained 
will be above the correct mean value (the other half being below it). 
In the limit of an infinite number of measurements the middle value is, 
by definition, the correct mean value.\footnote{Though strictly speaking
one should keep in mind that one can construct distributions that are
pathological enough to violate this.}

If each measurement is statistically independent, and with no assumptions
about the probability density function (PDF) or standard deviation of the
errors, there is a $50 \%$ chance that each measurement will be above (or
below) the true median value of the distribution.
So, if we perform $n$ measurements, the probability
that $k$ of them will be above (or below) the median is simply given by 
the binomial distribution,
\begin{equation}
P(k) = \frac{2^{-n}n!}{k!\ (n - k)!}\,.\label{binom}
\end{equation}
If we take $n$ measurements $M_n$ ordered from the smallest to the largest,
in such a way that $M_{i + 1} > M_i$, the probability of finding the median
between the measurements $M_i$ and $M_{i + 1}$ is again
\begin{equation}
P(i) = \frac{2^{-n}n!}{i!\ (n - i)!} \,, \label{probs}
\end{equation}
where we suppose $M_0 = - \infty$ and $M_{n + 1} = + \infty$. 

Given a Hubble diagram with the experimental results plotted against a
specific set of cosmological parameters,
\begin{equation}
\label{m_teo}
\mu _0(z) = 5 \log  \left(\frac{d_L(z, H_0, \Omega_m,
\Omega_\Lambda)}{{\rm {Mpc}}}\right) + 25\,,
\end{equation}
the relative likelihood associated with that set of parameters can be 
simply computed by counting the number of points above (or below) the 
expected curve and using eqn. (\ref{binom}).
When assuming Gaussian errors using a $\chi^2$ statistical treatment 
we benefit from the fact that doing so the
precision increases as $n^{-1/2}$, where $n$ is number of measurements.
Nonetheless, one can show \citep{Gott:2000mv} that with median statistics, and
relaxing assumptions (3) and (4), this result still holds.

It should be stressed that even though there is presently no evidence that
supernova luminosity errors are not Gaussian, calibrated light-curves are most
likely not Gaussian distributed. There are in fact indications that some
outliers are not well calibrated with the current methods of luminosity-curve
calibration \footnote{It is noteworthy that the use of different methods of
luminosity-curve normalisation carries an uncertainty of about
$\Delta m\sim 0.15$ magnitudes.}.
Median statistics are not as susceptible to these outliers as the more classic
$\chi^2$ analysis. \citet{Gott:2000mv} provide various
examples of how just one or very
few `fluke' data points could seriously distort a $\chi^2$ analysis,
and of why median statistics are much less vulnerable to such effects.

On the other hand, one should be wary of the fact that when
computing probabilities, the median statistic only
accounts for the number of experimental points above or below the expected
value. It does not differentiate between the various ways in which these
points could be distributed. Suppose that one has 10 magnitude \textit{versus}
redshift supernova measurements ordered by increasing redshift. A binomial
distribution associates a probability to the case where the first five
supernovae are brighter than expected and the last five fainter, equal to the
case where the first is brighter, the second fainter, the third brighter, the
fourth fainter and so forth until the tenth. These two cases should not be
indistinct since the first could turn out to be a terrible fit to the
data that happened to have exactly half of its points above and the other half
below it \footnote{It is a simple example to consider a horizontal
swarm of data points crossed by an almost vertical line through the middle,
leaving half the points in each side.}.

Note that this `sequence blindness' problem also exists, to some extent,
for the $\chi^2$ statistic. However, the crucial difference is that in this
case the error bars are know, which substantially attenuates the problem.

In order to improve the median statistic we will consider some adjustments to
its theoretical framework. Instead of just counting the number of data 
points above or
below the model prediction, we also take into account (a) the size of 
the largest contiguous
sequence found above or below the model prediction or (b) the number of
sequences obtained. With either of these we expect to more explicitly
account for the way in which the model under consideration intersects the
experimental data.

Let us be more specific about the modifications to the standard median 
statistic that we are presently proposing. Consider a random variable,
$X$, with a probability distribution with median $M$ and a number, $N$, 
of realizations of that variable. Assume for the sake of illustration the 
following result of $N=10$ realizations of the variable $X$
\begin{equation}
(-,+,+,-,+,+,+,-,-,+)
\end{equation}
where a $+(-)$ means that the particular realization ($X_i$ where $i$ can 
take any value between $1$ and $N$) of the variable $X$ is above
(below) the median of the distribution. In this particular case we can
see that there are $6$ sequences the largest of which has $3$ elements
$(+,+,+)$. We considered the probability, $P(k,N_s)$ of having at the
same time $k$ measurements above (or below) the median and a number, $N_s$, 
of sequences using a Monte Carlo simulation. We have also considered the 
probability $P(k,N_l)$ where in this case $N_l$ means the size of the
largest sequence. It is also important to refer that in our particular
application we consider that the measurements are naturally ordered by
increasing redshift.

Through a simple Monte Carlo simulation we compute the probabilities
$P(k,N_s)$ and $P(k,N_l)$ considering the number of data points
above or below the model prediction,
as well as the required sequence counting for each of
these two alternatives
(a) and (b). As we'll show below, these two alterations slightly 
improve the constraining process. We will also find that the performance of 
median statistic (and its modifications) strongly depends on the choice of 
the parameters being fitted for, due to reasons that will become apparent 
in the discussion.

\section{Results}

We now proceed to apply median and modified median
statistics to the current and SNAP supernova
datasets, and present confidence regions for the present day values of 
the cosmological parameters $\Omega_m$ and $\Omega_\Lambda$.
We also compare our results with the usual $\chi^2$ analysis.

\subsection{Standard median statistics}

\begin{figure}
\plottwo{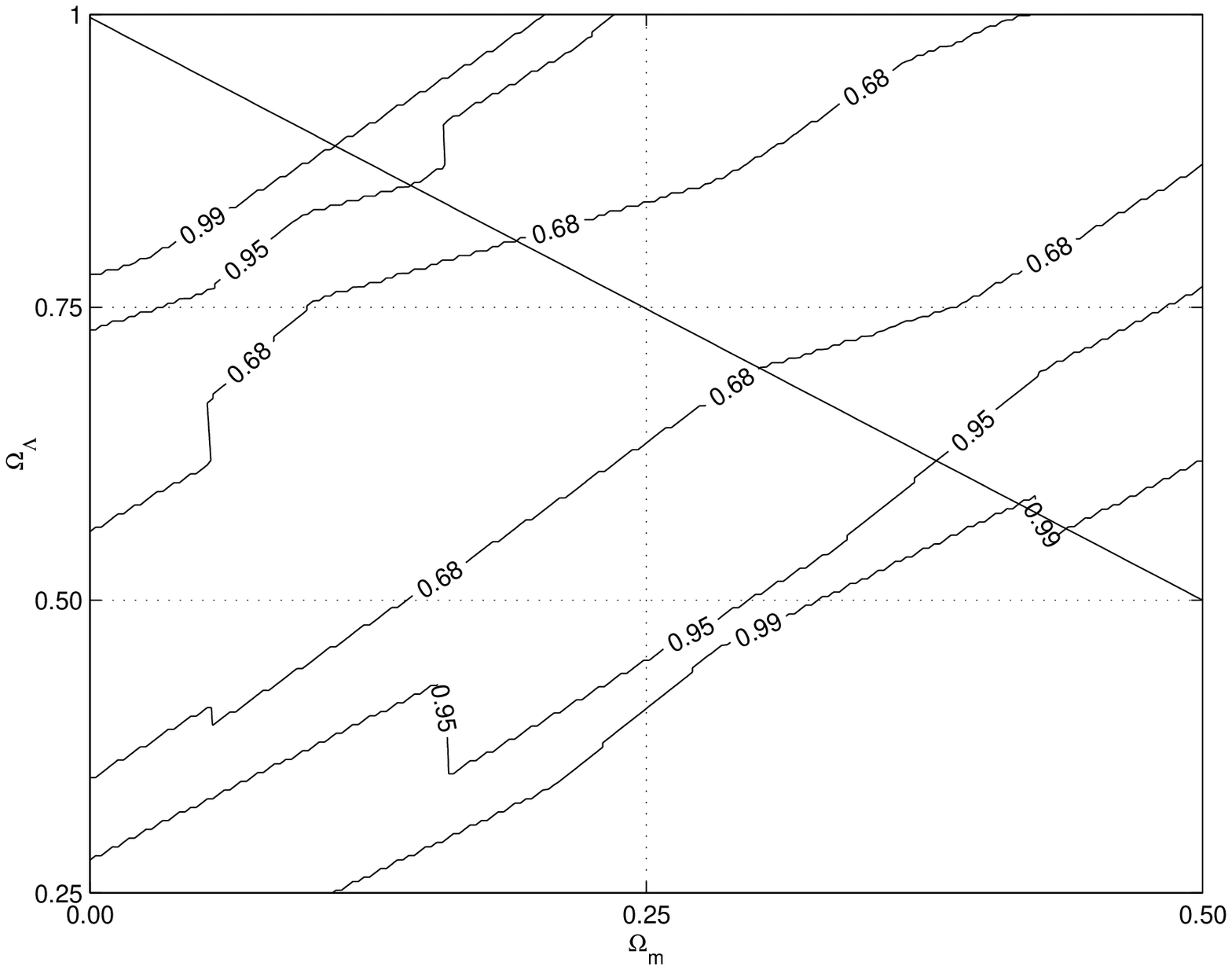}{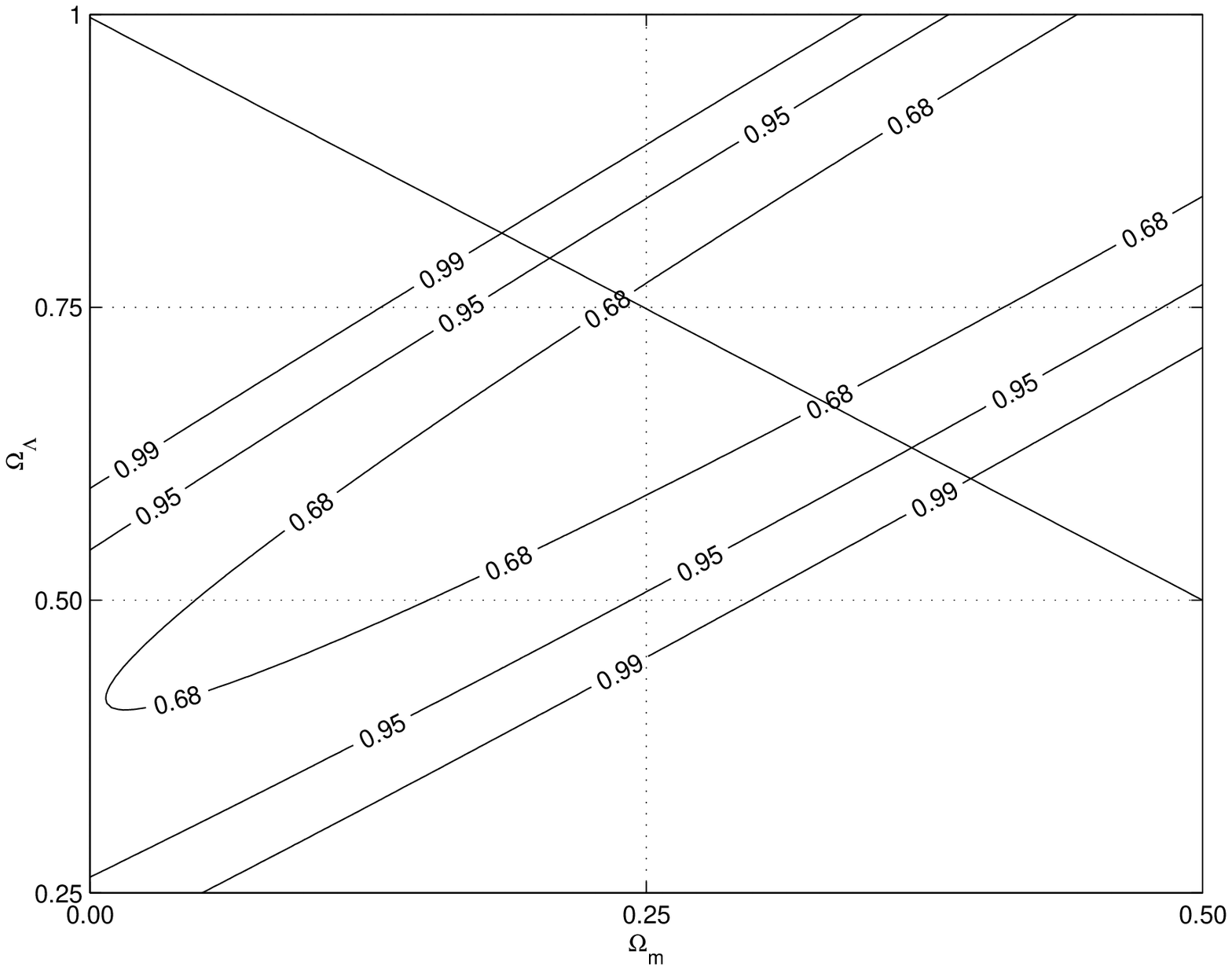}
\caption{The $68 \%$, $95 \%$ and $99 \% $ confidence regions
resulting from standard
median and $\chi^2$ statistics in the $\Omega_m-\Omega_{\Lambda}$ plane,
using the 92 supernovae from the combined SCP and HzST results. A value of
$H_0 = 65.2\ {\rm km\ s^{-1}\ Mpc^{-1}}$ was assumed for the Hubble
parameter (see main text). \label{fig1}}
\end{figure}

Results for the current supernova data set, using standard median and
$\chi^2$ statistics are shown in Fig. \ref{fig1} \footnote{This
updates the results of \citet{Gott:2000mv}, which only analysed
the results of the two supernova groups separately.}.
Here we assume the knowledge of the Hubble's constant, which we take to be
$H_0 = 65.2\ {\rm km\ s^{-1}\ Mpc^{-1}}$, in
agreement with \citet{Riess:1998cb}.

In a $\chi^2$ analysis it would be standard procedure to integrate over the
Hubble parameter,
\begin{equation}
P(\Omega_m, \Omega_\Lambda) = \int P(\Omega_m, \Omega_\Lambda, H_0) dH_0,
\end{equation}
but that is not the case with standard median statistics. In fact 
possibly the main
problem with this method is its inability to cope very well with a
multi-dimensional fit specially if it depends on several parameters. 
Clearly the Hubble parameter plays a very important part in this
analysis, and a good knowledge of it is necessary to obtain good results.
The factor $- 5\log H_0$, present in the distance modulus definition as an
additive constant, can move the zero-redshift
point up or down the magnitude scale leaving otherwise the curve 
unaffected. Recall that the supernovae distance scale depends on
the Large Magellanic Cloud's distance modulus, and indeed this is
the largest source of systematic error.

We note that given a sufficient number of local supernovae
it is possible to calibrate the value of $H_0$ since when $z \ll 1$ the
distance modulus is simply given by
\begin{equation}
\mu _0= 5\log \frac{cz}{H_0} + 25\,,
\end{equation}
independently of the other cosmological parameters. In this way the value
$H_0 = 65.2 \pm 1.3\ {\rm km\ s^{-1}\ Mpc^{-1}}$ (only the statistical error 
is included) was found by HzST \citep{Riess:1998cb} which is in agreement
with the HST Key Project result \citep{Freedman:1999wh}.

As expected, we find a somewhat larger confidence region in the case of
median statistics. Nevertheless, we can still exclude a
Universe with a vanishing cosmological constant with more than $99 \%$
confidence.
Similarly, the confidence region is above the 
$q_0 = \Omega_m/2-\Omega_\Lambda = 0$ line that separates an
accelerated expansion of the Universe from a decelerating one.

\begin{figure}
\plottwo{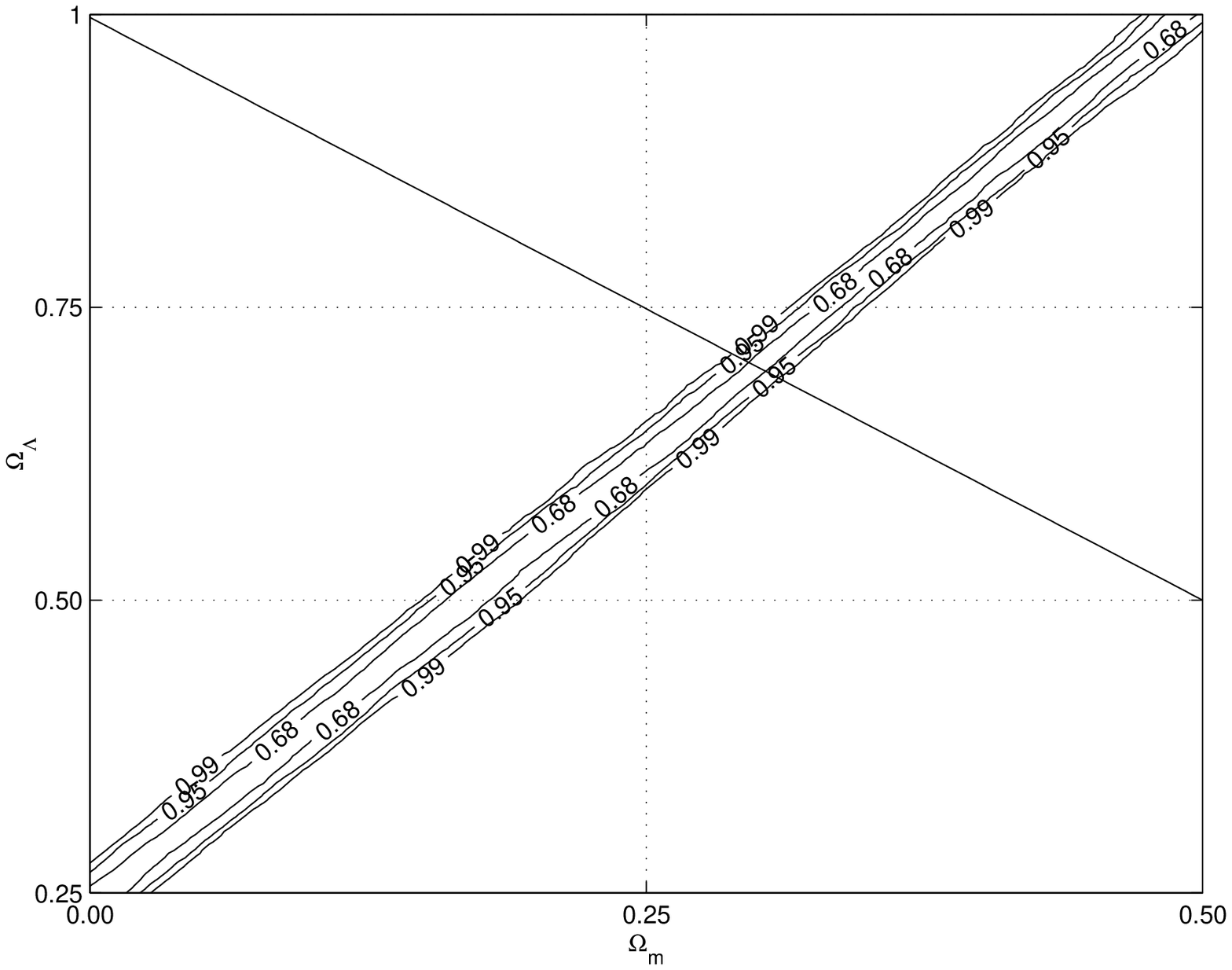}{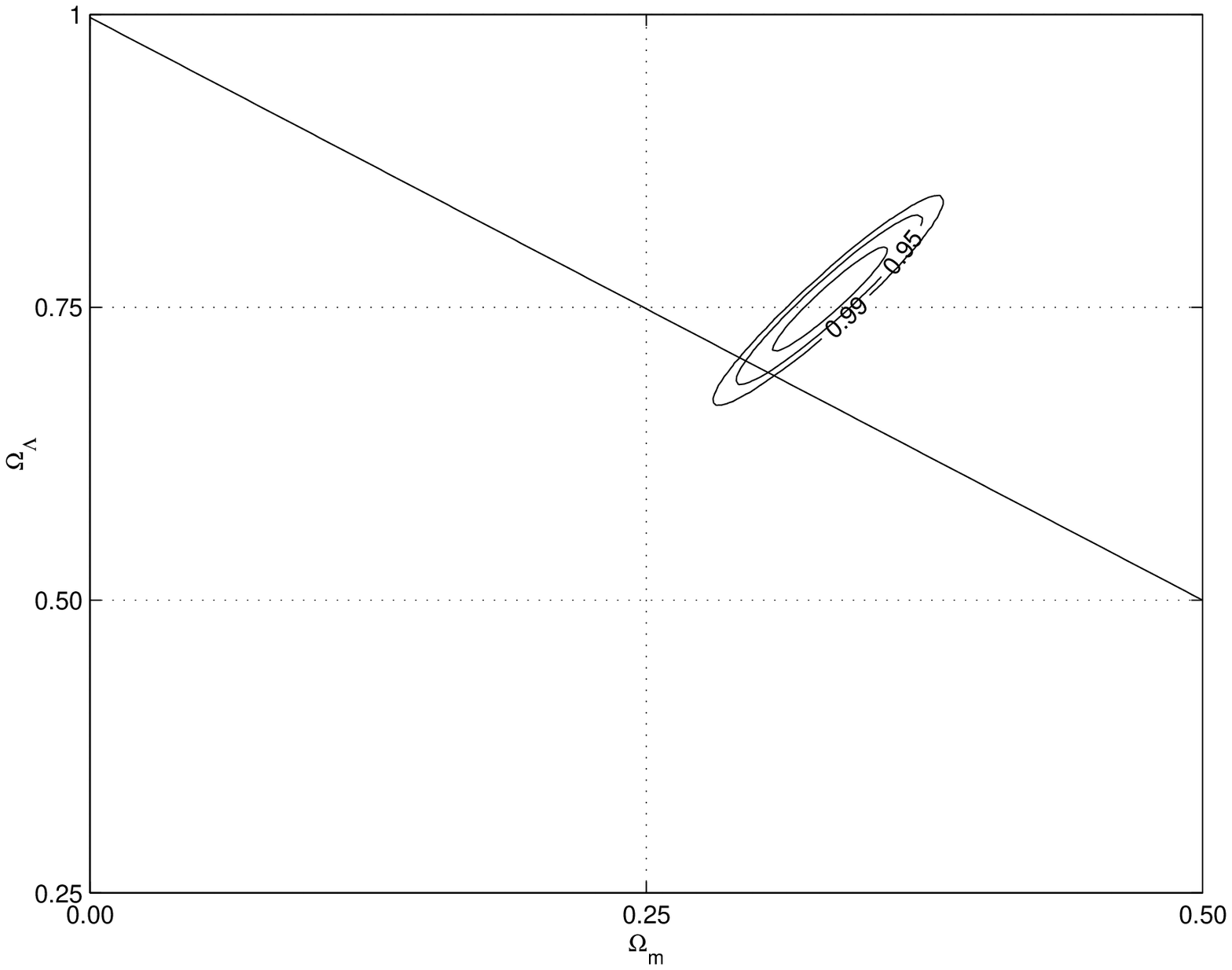}
\caption{The $68 \%$, $95 \%$ and $99 \% $ confidence
regions resulting from standard
median and $\chi^2$ statistics in the $\Omega_m-\Omega_{\Lambda}$ plane,
using the SNAP-class dataset. A value of
$H_0 = 65.2\ {\rm km\ s^{-1}\ Mpc^{-1}}$ was assumed for the Hubble
parameter (see main text). \label{fig2}}
\end{figure}

Fig. \ref{fig2} shows an analogous analysis but now using the SNAP
simulated results; again we assume the knowledge of the present day 
value of the Hubble parameter $H_0$. 
As expected, both statistics
can now accurately pin down a `degeneracy axis' but the error bars within 
this axis are considerably larger for median statistics.

\subsection{Modified median statistics}

So far we have done the analysis using the standard median statistics. 
We now consider the effect of the modifications which we described in
the in order to take into consideration
(a) the largest sequence or (b) the number of sequences obtained.
We shall see that 
these modifications allow us to slightly improve the constraints on 
the parameters being estimated.

\begin{figure}
\plottwo{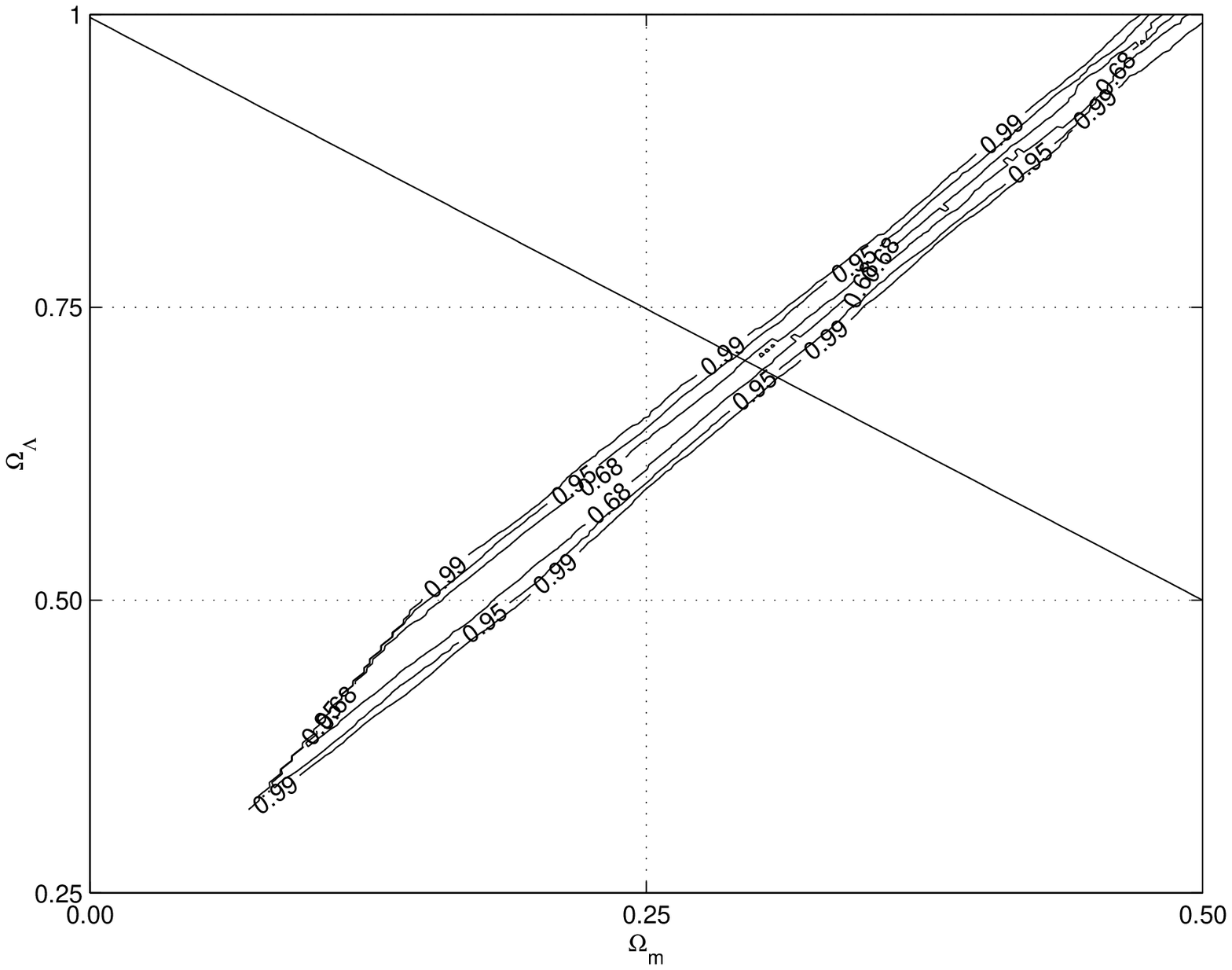}{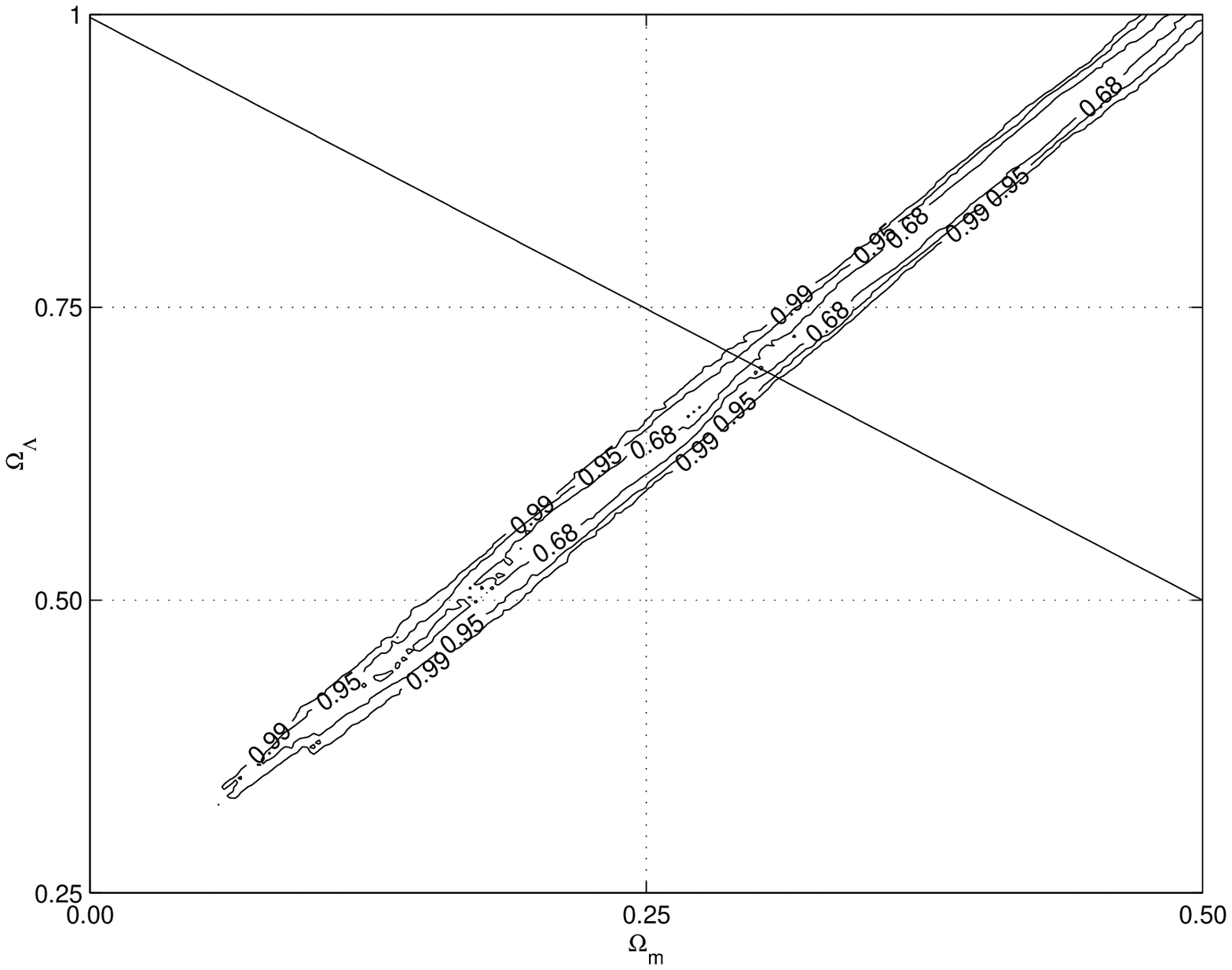}
\caption{The $68 \%$, $95 \%$ and $99 \% $ confidence
regions resulting from modified 
median statistics in the $\Omega_m-\Omega_{\Lambda}$ plane,
using the SNAP-class dataset. A value of
$H_0 = 65.2\ {\rm km\ s^{-1}\ Mpc^{-1}}$ was assumed for
the Hubble parameter (see main text).
The first plot shows the results obtained using median statistics modified to
account for the largest sequence, while in the second we consider the number
of sequences. \label{fig3}}
\end{figure}

Fig. \ref{fig3} shows the results of
analysing a SNAP-class mock dataset using median statistics modified
to include either of the two effects (a) or (b).

We can see that either modification seems to be 
more constraining than standard median statistics, as it reduces the
error bars within the above-mentioned `degeneracy axis'.
This was expected, for the reasons already pointed out above: a fit
where points alternate above and below a theoretical line should in
principle be better than one where there are long continuous sequences
either above or below it.

As one varies the present day values of the cosmological matter and 
vacuum densities, $\Omega_m$ and $\Omega_{\Lambda}$, the luminosity 
versus redshift curves also change. As a result of this tilting,
some of these curves will mostly be above (or below) 
the data points at high redshift.
A $\chi^2$ analysis
will immediately disfavour these models. On the other hand, in the case of
the standard median, many such models which can `compensate' for this by
having a fair amount of points at lower redshift below the data points will
still survive. However, if one accounts for the presence of sequences and
reduces the likelihood of any model where such sequences are found,
then one will be able to reduce the range of allowed models. We have verified 
that the performance of the modified median statistics upon integration 
over the Hubble parameter is significantly better than that of standard 
median statistic but still not competitive with the $\chi^2$ analysis.

We also note that the gain from either of the modifications is fairly similar.
Of course we could also implement them together if
desired. One would obtain a further (slight) improvement, at the expense
of having to deal with a somewhat more complicated statistic.

\subsection{The flat case}

Most inflationary models predict 
a flat Universe, and this seems to be confirmed by recent
results from CMB anisotropy measurements \citep{Jaffe:2000tx}.
Using this prior, the precision of the fits is quite substantially
increased, as we'll be fitting for a single parameter. 
This can already be seen in Figs.
\ref{fig1}\textendash\ref{fig3} where the diagonal line that 
intersects the confidence regions represents the combinations of the 
cosmological parameters $\Omega_m$ and $\Omega_\Lambda$ which correspond 
to a flat universe.

In the case of a flat universe the modifications we made to median 
statistics do not significantly improve 
the constraints on $\Omega_m-\Omega_\Lambda$ plane obtained using 
the conventional median analysis and so we'll restrict ourselves to the 
standard case.

\begin{figure}
\plottwo{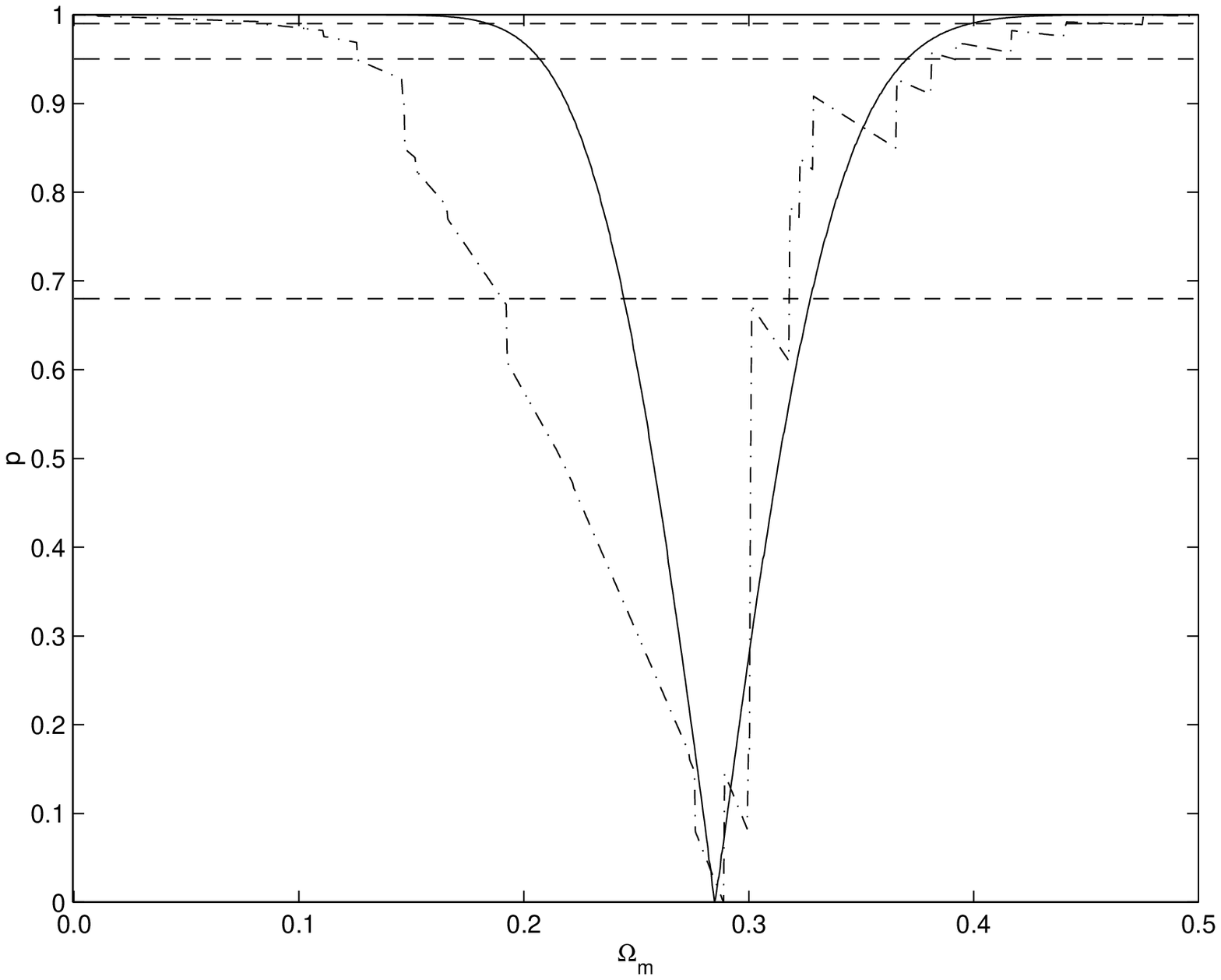}{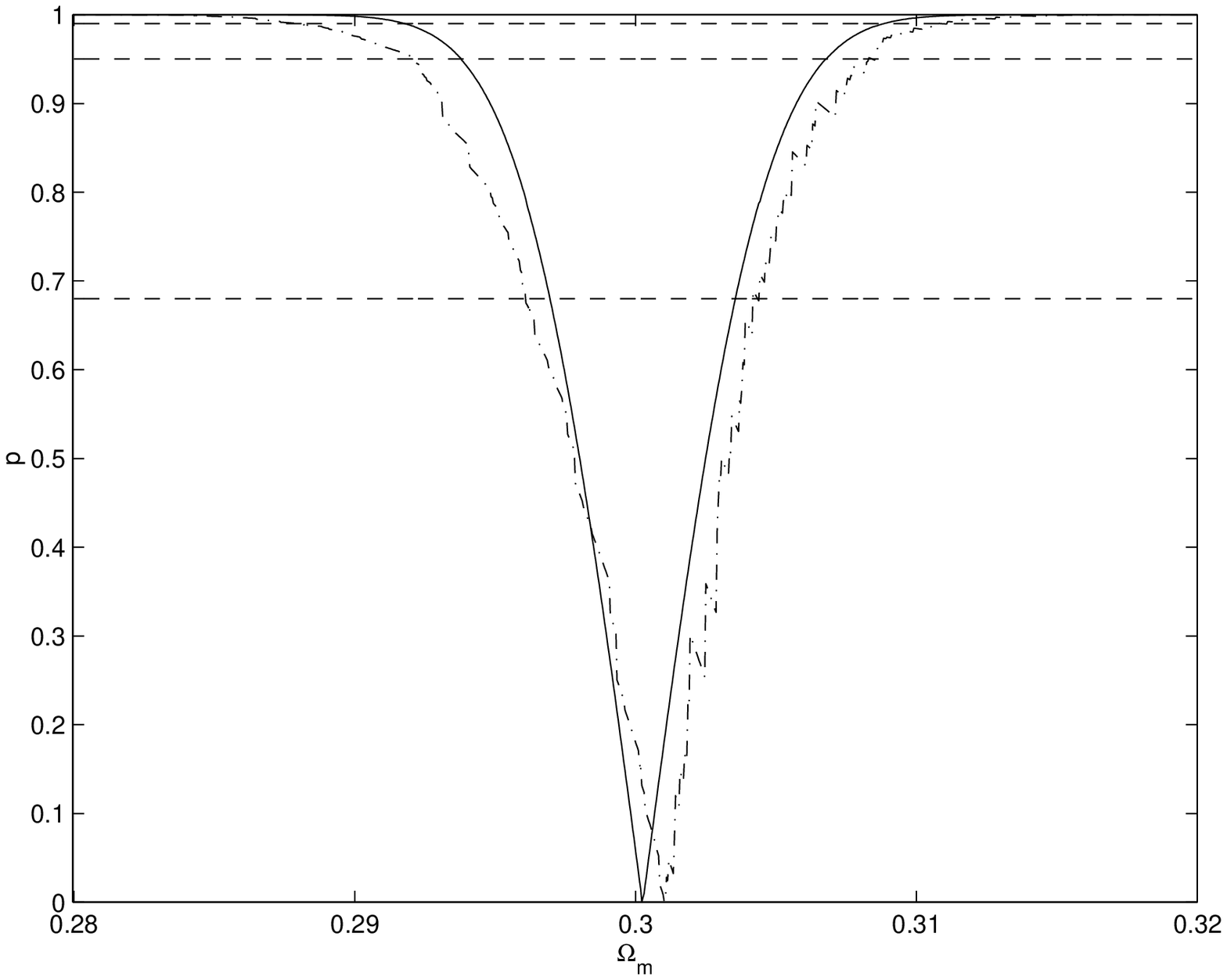}
\caption{Marginal distributions for $\Omega_m$ obtained using the standard 
median (dot-dashed) and $\chi^2$ (solid line) statistics, 
assuming a flat Universe, for the current and
the SNAP-class datasets. The $68 \%$, $95 \%$ and 
$99 \%$ confidence limits are where the curves drop below the 
(dashed) lines of constant $p$ (0.68,0.95 and 0.99 respectively). 
Note that a value of $H_0 = 65.2\ {\rm km\ s^{-1}\ Mpc^{-1}}$ was 
assumed for the Hubble parameter. \label{fig4}}
\end{figure}

In Fig. \ref{fig4} we show results obtained for both datasets through the
methods previously presented, assuming a flat universe prior using the
conventional median statistic and the $\chi^2$ statistic.
We present these results in a more convenient form
in table \ref{median_1D_results}.

For a flat Universe we obtain $\Omega_m = 0.29^{+0.03}_{-0.10}$ (not including 
the uncertainty in the Hubble parameter) using currently available 
supernovae and simple median statistics.
For these 92 supernovae the median statistic is marginally less competitive 
than the $\chi^2$ statistic and one notices that the median results are 
slightly skewed towards low values of $\Omega_m$. Nevertheless the $90 \%$ 
confidence upper limit for $\Omega_m$ is slightly lower in the case of 
median statistics and the confidence intervals still overlap nicely. 

Clearly the SNAP results are expected to significantly improve the
constraints on the energy density of the Universe. Note that the results
obtained with median statistics are almost indistinguishable from those 
obtained with the standard $\chi^2$ analysis in this case.

This analysis clearly show that the median and the
standard $\chi^2$ statistics have quite similar performances if the 
assumptions about a Gaussian distribution for the errors and the 
estimate of the standard deviations are correct. If this is not the
case, then obviously the results obtained with the median statistic are
the more reliable ones.

Indeed, one could use this information to reverse the argument
and use both statistics together as a test on the assumptions being made
on the data. If in the case of SNAP the median and $\chi^2$ statistics
do not agree, then this indicate that either the errors do not have a 
Gaussian distribution or that one is somehow underestimating
the statistical and/or systematic errors.

\begin{table}
\caption{\label{median_1D_results}Values of $\Omega_m $ obtained
through median statistics for supernova data
(both the current and the SNAP datasets) with a flat universe prior.
Results for the usual $\chi^2$ statistic are also shown for comparison. 
Here a value of $H_0 = 65.2\ {\rm km\ s^{-1}\ Mpc^{-1}}$ was 
assumed for the Hubble parameter}
\begin{tabular}{|c|c|c|}
Statistic               & Current data set      & SNAP data set       \\
\hline
Median & \( 0.29^{+0.03}_{-0.10} \) & \( 0.301^{+0.004}_{-0.005} \) \\ 
Chi Squared  & \( 0.28^{+0.05}_{-0.04} \) & \( 0.300^{+0.004}_{-0.003} \) \\
\end{tabular}
\end{table}

\section{Conclusions}

We have discussed standard and modified median statistics in the context
of current and forthcoming type Ia supernova data sets. The purpose of the
modifications is to reduce some of its weaknesses, mainly by accounting for
the number and size of sequences of consecutive points above or
below the median. We found that in some circumstances the performances of the
median and $\chi^2$ statistics can be comparable, and if used together they
provide a simple test on the assumptions being made on the data.

The main problem with the standard median statistics analysis is its 
inability to cope well with a multi-dimensional fit specially if it 
is dependent on several parameters. This is due to the very simple 
assumptions it makes. On the other hand, when confronted 
with a single parameter to fit, the ensuing results can be of similar 
precision to the ones obtained with a $\chi^2$ analysis.

Another advantage of median statistics is that it is an analysis method
which is extremely easy to implement. So even in the cases where
it is not expected to produce results as constraining as a $\chi^2$ analysis,
it can be used to complement it or to provide fast fits in the early stages
of the analysis, notably if one is still trying to gather supporting
evidence for the use of stronger hypothesis about the dataset.
Using median statistics we no longer have to suppose
that the errors follow a Gaussian distribution with known standard deviation,
and can therefore have greater confidence in the parameter ranges obtained.
This is an important concern for the particular case of type Ia supernovae:
recall that when studying them we are considering renormalised light curves,
and that the calibration data set (of nearby supernovae) is smaller than the
main data set (of distant supernovae). The median statistic is also less 
sensitive to systematic effects such as weak lensing.

We have studied some simple means by which to improve median statistics,
namely accounting for the size of the largest sequence or the number of 
sequences present in the dataset above or below the model prediction, 
Our adjustments did provide some improvement, even if in a 
our confidence regions are still larger than that obtained
from a $\chi^2$ study. We note however that other more `baroque'
modifications are certainly conceivable. Other possibilities, which certainly
deserve further work, are to study how similar procedures could be 
used to improve the standard $\chi^2$ analysis (which is sequence blind), 
and to apply median statistics to other cosmological data sets, 
most notably the cosmic microwave background. We shall report on these 
issues in a forthcoming publication.

\acknowledgments

C.M. is funded by FCT (Portugal), under grant no. FMRH/BPD/1600/2000.
We are grateful to Alessandro Melchiorri, Gra\c ca Rocha and Paul Shellard
for useful comments and suggestions.

\clearpage

\end{document}